\documentclass[11pt]{article}
\usepackage[margin=1.0in]{geometry}
\usepackage{graphicx}
\usepackage{epsfig}
\usepackage{amssymb,amsfonts,amsmath,amsthm}
\usepackage{mathrsfs}
\usepackage{epstopdf}
\usepackage{hyperref,color}
\usepackage{natbib}
\usepackage{setspace}

\pdfoutput=1

\begin{document}

\title{Time travelling in emergent spacetime}
\author{Christian W\"uthrich\thanks{I am grateful to the editors for their kind invitation and to Hajnal Andr\'eka, Stefano Furlan, Niels Linnemann, Istv\'an N\'emeti and an anonymous referee for their comments on earlier versions of this paper and for discussions. I am also grateful to Hajnal Andr\'eka and Istv\'an N\'emeti for their collaboration on earlier projects. But most of all, I am honoured by their friendship.}}
\date{21 June 2019}
\maketitle

\noindent
{\small For Judit Madar\'asz and Gergely Sz\'ekely (eds.), {\em Hajnal Andr\'eka and Istv\'an N\'emeti on the Unity of Science: From Computing to Relativity Theory Through Algebraic Logic}, Springer.}

\begin{abstract}\noindent
Most approaches to quantum gravity suggest that relativistic spacetime is not fundamental, but instead emerges from some non-spatiotemporal structure. This paper investigates the implications of this suggestion for the possibility of time travel in the sense of the existence of closed timelike curves in some relativistic spacetimes. In short, will quantum gravity reverse or strengthen general relativity's verdict that time travel is possible?
\end{abstract}

\section{Introduction}\label{sec:intro}

General relativity (GR), our currently best theory of gravity and of spacetime, permits time travel into one's own past in the sense that it contains models of spacetime with \textit{closed timelike curves}, i.e., worldlines potentially traced out by matter in spacetime, which intersect themselves. If a particle follows such a closed worldline, it returns not only to its earlier position in space---which is common enough---, but in space\textit{time}, i.e., also to its earlier position in time. An early example of such a spacetime is what has become known as \textit{G\"odel spacetime} \citep{god49}, but in fact there are innumerably many such solutions in GR.\footnote{For a recent review, see \citet{smewut11}.} Should we thus conclude that time travel is, in fact, physically possible, i.e., in accord with the laws of nature? 

We should not, as there are good reasons to think that, despite its phenomenal empirical success, GR is not the last word on gravity and on the fundamental structure of what plays the role of spacetime: GR assumes that matter has essentially classical properties, e.g., by having a determinate spatiotemporal distribution. But of course we have learned from quantum physics that matter degrees of freedom behave rather differently. Thus, at a minimum, GR ought to be replaced by a theory which can accommodate the quantum nature of matter. It is for this simple but conclusive reason that we need a quantum theory of gravity.\footnote{And much less for a whole list of other reasons routinely given in the literature, and critically discussed by \citet{hugcal01,wut05,mat06}.} No such theory yet exists in its fully articulated and empirically confirmed form, but candidate theories are string theory, loop quantum gravity, causal set theory, and many more. 

Thus arises the question of whether these candidates for a more fundamental theory of gravity permit time travel in the same or a similar sense as does GR. In fact, there are two ways in which a quantum theory of gravity might do so. First, it may permit time travel by admitting models which contain the (analogue of) closed timelike curves. In this case, time travel would accord to the laws of nature stipulated by that theory. This would straightforwardly licence time travel's physical possibility. Second, although that theory itself may prohibit time travel in this same sense, it could allow for the emergent relativistic spacetime---which well approximates the fundamental structure at some scale---to contain closed timelike curves. Although the fundamental theory would then remain inhospitable to time travel itself, it would tolerate the possibility of time travel at some other, less fundamental, scale. It is this possibility in particular that I wish to explore in this article.

After settings things up in Section \ref{sec:classical}, I will introduce four theories of quantum gravity in Section \ref{sec:qg}: semi-classical quantum gravity, causal set theory, loop quantum gravity, and string theory, and discuss the possibility of time travel directly in those theories. In Section \ref{sec:emergent}, I will turn to the second possibility, viz., that these theories themselves disallow time travel, but fail to prevent it at the emergent level. Conclusions follow in Section \ref{sec:conc}. 

\section{Global hyperbolicity and energy conditions}\label{sec:classical}

Does a theory succeeding to GR include or exclude closed timelike curves and similar causal pathologies inside the bounds of what it deems physically possible? Since despite valiant efforts, no quantum theory of gravity has been fully articulated, let alone empirically confirmed, our discussion must remain preliminary and speculative. Still, from considering candidate theories and their presumed verdict on the question, we hope to glean intimations of an answer and at least start to survey the dialectical landscape of possibilities. 

There are at least three ways in which quantum gravity may prejudge the case for or against the possibility of time travel. First, it may rule it out {\em by fiat} by imposing global hyperbolicity or a kindred mathematical condition. Such an imposition may be metaphysically motivated to rule out causal pathologies, or it may be occasioned by the pragmatic desire to apply a particular mathematical apparatus, which requires the condition. This {\em a priori} restriction to causally benign structures may, of course, eventually be justified {\em a posteriori} by the empirical success of the theory. 

Second, it may be the case that although no such condition is demanded at the outset, it can be derived from the resources of the theory itself. In particular, time travel may be ruled out as a consequence of well-justified assumptions concerning what is physically reasonable or even possible.  In a situation like this, it might appear as if time travel is ruled out on physical grounds, and that causality-violating spacetimes ought to be deemed unphysical artefacts of the mathematical formalism of overly permissive GR.\footnote{It certainly appeared so to me when Smeenk and I wrote \citet[\S8]{smewut11}.} Although suggestive, we will see in \S\ref{sec:emergent} that this may not follow.

The third possibility is that we find closed timelike curves (or an analogous feature) prevalent in the more fundamental theory of quantum gravity, or at least in its physical applications. This outcome would suggest (though perhaps again not entail, see \S\ref{sec:emergent} below) that the intriguing possibility of violations of causality, first encountered in GR, may remain in quantum gravity.

Before turning to approaches to quantum gravity in the next section, the remainder of this section (\S\ref{ssec:classical}) offers a brief discussion at the classical level of two notions central for the possibility of closed timelike curves and thus of time travel: global hyperbolicity and so-called `energy conditions'. 

\subsection{At the classical level}\label{ssec:classical}

Let me start by fixing some terminology. A relativistic spacetime $\langle M, g_{ab}\rangle$ described by a four-dimensional, differentiable manifold $M$ with a metric $g_{ab}$ with Lorentz signature is \textit{time orientable} just in case it permits a globally consistent time direction in the form of an everywhere defined continuous timelike vector field. As a temporal direction is picked as the `future', the time orientation of such a spacetime is thereby determined. A \textit{worldline} is a continuous timelike curve whose orientation agrees with the time orientation of the spacetime in which it is contained. A \textit{closed timelike curve} is a closed worldline. 

The existence of closed timelike curves in a spacetime mark the violation of a so-called `causality condition'. It turns out that there is whole hierarchy of stronger and stronger causality conditions (\citealt[\S\S6.4-6.6]{hawell73}; \citealt[\S\S8.2-8.3]{wal84}). The weakest condition requires that there are no closed timelike curves. The strongest condition demands that the spacetime be `globally hyperbolic'. Thus, if a spacetime is globally hyperbolic, it does not contain closed timelike curves. 

Let us unpack the notion of global hyperbolicity. A spacelike hypersurface $\Sigma \subset M$ with no edges is called a \textit{global time slice}. If such a global time slice $\Sigma$ is \textit{achronal}, i.e., if it is not intersected more than once by any future-directed causal curve, it is called a \textit{partial Cauchy surface}. The \textit{future domain of dependence} $D^+(\Sigma)$ of a partial Cauchy surface $\Sigma$ is the set of events $p\in M$ such that every past inextendible causal curve through $p$ intersects $\Sigma$. The \textit{past domain of dependence} $D^-(\Sigma)$ is defined analogously. A partial Cauchy surface $\Sigma$ is a \textit{Cauchy surface} just in case the total domain of dependence $D^+(\Sigma) \cup D^-(\Sigma)$ is $M$. A spacetime $\langle M, g_{ab}\rangle$ which admits a Cauchy surface is said to be \textit{globally hyperbolic}. 

The future domain of dependence of a global time slice $\Sigma$ is of interest because it characterises the set of events for which any signal or information which reaches them must have passed through $\Sigma$. Thus, assuming that signals and information cannot travel faster than the speed of light, conditions on $\Sigma$ should determine the complete state at $p$---assuming deterministic dynamics.\footnote{This expectation is confirmed, e.g., for physical fields in curved spacetimes, which propagate in accordance to hyperbolic wave equations \citep[Ch.\ 10]{wal84}.} Similarly, the conditions on $\Sigma$ should determine the state at any event in the past domain of dependence. In the case of a globally hyperbolic spacetime, therefore, any event at all in the spacetime is similarly determined by the conditions on $\Sigma$. Consequently, there cannot be (among other things) closed timelike curves in such a spacetime: regardless of whether or not these closed timelike curves intersect $\Sigma$, $\Sigma$ would not be a Cauchy surface, and so the spacetime would not be globally hyperbolic. Global hyperbolicity or the existence of closed timelike curves are global properties of a spacetime in the sense that, although properties ascribable to spacetimes, they are not possessed by individual events and do not supervene on any such local properties.

Before we get to quantum gravity, it should be noted that it would be surprising if moving beyond GR would mean relapsing into imposing global, non-dynamical constraints on spacetime structure such as prohibiting the existence of closed timelike curves by fiat, as it appears as if GR owes its success precisely to abandoning such constraints. We should expect one kind of constraint, however, to restrict the models of classical GR: energy conditions. These are universally valid (but local) constraints on the matter sector of the theory and capture the thought that not just any stress-energy tensor $T_{ab}$ can adequately represent the physical matter content of the universe. Thus, they express general conditions which any matter or non-gravitational field is required to satisfy in order to qualify as `physical'. Through the Einstein equation, 
\begin{equation}\label{eq:efe}
G_{ab} = 8\pi T_{ab},
\end{equation}
where the \textit{Einstein tensor} $G_{ab} = G_{ab} [g_{ab}]$ is constructed from the spacetime metric $g_{ab}$ and its first and second derivatives, and Newton's constant $G$ and the speed of light $c$ are set to 1, that matter content is related to the geometry of the spacetime. Just like the metric, the Einstein tensor is defined on a four-dimensional pseudo-Riemannian manifold $M$  and describes the curvature of the spacetime $\langle M, g_{ab}\rangle$. Importantly, the (classical) energy conditions are defined in tangent spaces and so obtain locally, i.e., point-wise. Since these conditions hold only strictly locally, they do not have the power to rule out causal pathologies such as closed timelike curves, which are global (or at least `regional') properties of a spacetime in that they are topological features of a spacetime that can only be exemplified in at least a region of spacetime. 

As it turns out, these point-wise energy conditions can only be satisfied for types of `classical' matter; they all fail for quantum fields (due to arbitrarily negative expectation values of energy densities of quantum fields at any point). Hence, the classical conditions have been relaxed to `non-local' energy conditions, which hold in extended regions of spacetime, rather than at single events. Thus, they could at least potentially disqualify spacetimes with closed timelike curves as unphysical. Although the final verdict is out, it seems, however, that this hope will not be borne out.\footnote{For a discussion of this point, see \citet[\S7]{smewut11}; for a primer on energy conditions, see \citet{cur17}.}

\section{Theories of quantum gravity}\label{sec:qg}

This section will introduce four approaches to quantum gravity and discuss the viability of time travel in each of them: semi-classical quantum gravity (\S\ref{ssec:semi}), causal set theory (\S\ref{ssec:cst}), loop quantum gravity (\S\ref{ssec:lqg}), and string theory (\S\ref{ssec:st}). 

\subsection{Semi-classical quantum gravity}\label{ssec:semi}

The research programme of quantum field theory on curved spacetime offers a first stab at a quantum theory of gravity. Although mathematically demanding, the approach is physically simple: take a classical relativistic spacetime and treat it as a fixed background for quantum fields. For a linear field $\phi$ defined over globally hyperbolic spacetimes $\langle M, g_{ab}\rangle$, there is a mathematically rigorous and physically well-behaved procedure for writing down an algebra $\mathfrak{A}(\langle M, g_{ab}\rangle)$ of observables \citep[Ch.\ 4]{wal94}. For more general fields, however, semi-classical quantum gravity may not be well-behaved, and the procedure cannot be applied to non-globally hyperbolic spacetimes.\footnote{For a recent---and optimistic---review, see \citet{ver12}.} If the approach demands global hyperbolicity, it cannot accommodate time travel on closed timelike curves. Since the spacetime is already set in place---and fixed---, there is also no option of emergent time travel under the assumption of global hyperbolicity. This will play out rather differently in loop quantum gravity (see below). 

However, the most severe limitation of the approach is that the spacetime structure is assumed to be fixed. This stands in obvious tension with the insight in GR that the spacetime geometry not only acts upon the matter content of the world, but is also acted upon by it. Thus, spacetime geometry is dynamical, and one must countenance the `backreaction' of the matter field on the metric. The most basic way to construct a quantum theory of gravity which does this is to combine classical relativistic spacetime geometry---the left-hand side of (\ref{eq:efe})---with an account of quantum matter which will determine the right-hand side of (\ref{eq:efe}). The quantum matter fields, described by an appropriate quantum field theory (QFT), propagate in a classical spacetime. The backreaction of the matter fields on the spacetime geometry is computed through the \textit{semi-classical Einstein field equation}:
\begin{equation}\label{eq:semi}
G_{ab} = 8 \pi \langle \psi |\hat{T}_{ab} |\psi\rangle,
\end{equation}
where $\langle \psi |\hat{T}_{ab} |\psi\rangle$ is the expectation value of the stress-energy tensor of the quantum fields (which now is of course an operator) in a (physically reasonable) state $|\psi\rangle$. Semi-classical quantum gravity is a quantum theory of gravity as defined above: it combines gravity---in the form of spacetime curvature---with a quantum theory of matter. In general, semi-classical quantum gravity is expected to offer a valid extension of GR for some relatively simple cases when quantum and gravitational effects are not too strong \citep[\S2]{wut19a}. 

Does semi-classical quantum gravity permit time travel? As the only difference between the fully classical equation (\ref{eq:efe}) and the semi-classical one (\ref{eq:semi}) is in the description of the matter on the right-hand side, the relevant issue is whether the quantum nature of matter is less, equally, or more constraining on the spacetime geometry on the left-hand side than is classical matter. 

As mentioned above, the most direct way in which it is \textit{less} constraining and so more permissive of time travel than classical matter is by violating the energy conditions believed to hold for classical matter. However, the expectation value $\langle \psi |\hat{T}_{ab} |\psi\rangle$ may also act in ways which are \textit{more} constraining than classical matter. For instance, \citet{haw92} argued that since $\langle \psi |\hat{T}_{ab} |\psi\rangle$ appears to diverge on or near the boundary to the region of spacetime containing closed timelike curves (assuming there were none `before'),\footnote{These boundaries are so-called `future Cauchy horizons', i.e., boundaries of future domains of dependence of global time slides, where these domains are defined as those regions such that  every past inextendible causal or timelike through any event in the region intersects the global time slice.} it effectively `cuts off' the region of spacetime with the causal pathologies, rendering it inaccessible from the causally well-behaved domain and thus effectively protecting `chronology'. Hawking took the divergence of the expectation value of the energy-momentum tensor as the Cauchy horizon is approached and thus as closed timelike curves are `about to form' to strongly support his `chronology protection conjecture', according to which the ``\textit{laws of physics do not allow the appearance of closed timelike curves}'' \citep[603, emphasis in original]{haw92}. Stated in this way, given the pervasive existence of closed timelike curves in relativistic spacetimes, there thus seems to be little reason to think that the chronology protection conjecture is true in semi-classical quantum gravity, and no reason at all to accept it in the context of GR. 

If successful, Hawking's argument might well only establish that the region with the closed timelike curves is beyond the reach of physical denizens of the causally well-behaved region on this side of the Cauchy horizon as they would have to pass through a wall of arbitrarily high energy density in order to be able to travel along closed timelike curves. But these curves might still exist beyond the Cauchy horizon in an inaccessible region of spacetime, and in fact could be taken advantage of by would-be time travellers on the far side of the Cauchy horizon. In this case, the laws of physics would not prevent the \textit{existence} of closed timelike curves, though perhaps their \textit{accessibility}. 

However, it is not clear whether Hawking's argument succeeds in the first place. A theorem due to \citet{kayeal97} establishes that the expectation value of the energy-momentum tensor for a scalar field is not everywhere well-defined on compactly generated Cauchy horizons. The authors suggest that this result may be taken as further support of Hawking's chronology protection conjecture in that it suggests that the Cauchy horizon cordons off the region with closed timelike curves. However, the result can just as well be taken to indicate that semi-classical quantum gravity is simply no longer valid at the horizon\footnote{See for example \citet{vis03}, as well as the discussion in \citet[\S5]{earsmewut}.} and that therefore, Hawking's argument fails, at least if based on solely on semi-classical quantum gravity. Thus, only a more fundamental theory of quantum gravity can deliver a final verdict on the matter. The prospect of probing more deeply to see whether chronology protection obtains motivates not only the present inquiry, but---as should become clear---also promises to shed light on the nature of quantum gravity itself.

\subsection{Causal set theory}\label{ssec:cst}

Just as the research programs introduced in \S\ref{ssec:lqg} and \S\ref{ssec:st}, causal set theory aims at offering a `full' quantum theory of gravity, i.e., a theory in which also the gravitational sector is subjected to a quantum treatment. It is motivated by a result in classical GR, which shows that at least for an important class of relativistic spacetimes, the causal structure determines the metric structure of the spacetime up to a conformal factor.\footnote{This is a paraphrase of a theorem due to \citet{mal77}. More precisely, the theorem states that for any two `distinguishing' (and temporally oriented) spacetimes $\langle M, g_{ab}\rangle$ and $\langle M', g'_{ab}\rangle$, a causal isomorphism $\phi: M \rightarrow M'$ is a smooth conformal isometry. A bijection $\phi: M \rightarrow M'$ is a \textit{causal isomorphism} just in case for all $p, q \in M$, $p$ is in the chronological past of $q$ if and only if $\phi(p)$ is in the chronological past of $\phi(q)$. A spacetime $\langle M, g_{ab}\rangle$ is \textit{distinguishing} just in case for all $p, q \in M$, if the chronological past of $p$ is identical to the chronological past of $q$, then $p=q$, and if the chronological future of $p$ is identical to the chronological future of $q$, then $p=q$. A causal isomorphism $\phi$ is a conformal isometry just in case it is a diffeomorphism and there exists a conformal factor $\Omega: M' \rightarrow \mathbb{R}$ such that $\phi_\ast (g_{ab}) = \Omega^2 g'_{ab}$ with $\Omega \neq 0$.} This result is interpreted to suggest that the causal structure of a (causally well-behaved) spacetime contains almost the full information concerning its geometry; in fact, all but some information about local `size'. In the causal set theory programme, this missing `size' information is naturally supplied by the number of discrete `atoms' of spacetime contained in any region. In slogan form, causal set theory assumes spacetime to be causal structure plus number. Accordingly, the fundamental structures postulated by causal set theory---the `causal sets'---are discrete sets of elementary events, which are partially ordered by a relation of causal precedence or of causal connectibility. 

As it stands today, causal set theory frames a promising research programme but is still a long way from offering a complete quantum theory of gravity. The promise of the research programme remains unfulfilled in three ways. First, merely requiring the fundamental structure of our world to be a discrete, partially ordered set falls way short of constraining the boundless possible combinations of such structures to serious candidates with a promise to reproduce our physical world: there are just too many discrete partial orders, almost all of which do not resemble our universe. How can one identify the `physical sector' of the theory? The most popular strategy to taming the unruly possibilities is by imposing additional constraints; in particular, advocates of causal set theory favour imposing dynamical laws in response to the problem (e.g.\ the classical sequential growth dynamics proposed in \citealt{ridsor99}). 

Second, even the successful resolution of this trouble would at best result in a purely classical theory: neither does the state space have the structure of a vector space, nor is there anything quantum about the dynamics. If the theory is truly to incorporate the quantum nature of matter, then causal set theory as it stands can at best be a stepping stone toward a full quantum theory of gravity. Third, causal set theory suffers from the same affliction as all other approaches to quantum gravity: a full understanding of the relationship between the fundamental physics postulated and the emergent relativistic spacetime with its dynamics between spacetime and matter as encoded in the Einstein field equation remains elusive.

Whatever the eventual resolution of these challenges may look like, what does the present state of the theory suggest regarding the possibility of time travel? In special-relativistic theories, the causal structure of spacetime is expressed by the usual and well-behaved lightcones of Minkowski spacetime. In Minkowski spacetime, causal precedence thus merely \textit{partially} orders events, as spacelike related events do not stand in this relation. Since causal set theory also permits `spacelike separated' events, the ordering is equally merely partial. 

For our present purposes what matters, however, is that the ordering is also no weaker than partial. This means, in particular, that it is not a mere \textit{pre-order}, i.e., a reflexive and transitive order, which is not, in general, antisymmetric.\footnote{A binary relation $R$ on a domain $D$ is \textit{antisymmetric} just in case for all $x, y \in D$, if $Rxy$ and $Ryx$, then $x=y$.} The demand that the causal relation be antisymmetric (and so not a mere pre-order) thus precludes the possibility of causal loops of the form of cycles containing numerically distinct events $a$ and $b$ such that $a$ causally precedes $b$ and $b$ causally precedes $a$, as was possible in spacetimes in GR which contain closed timelike curves. In other words, causal set theory prohibits, in its central axiom, that the fundamental structure accommodates what would be the natural analogue of closed timelike curves in causal set theory, i.e., closed chains of events connected by the relation of causal connectibility. 

This choice simplifies the technical demands of the approach, as well as its metaphysics (\citealt{wut12c} and \citealt[Ch.\ 3]{hugwut}), but it imperils causal set theory's capacity to give rise to relativistic spacetime models which do include closed timelike curves. Although physicists are generally happy to give up non-globally hyperbolic models of GR, a theory's inability to reproduce that sector of GR may turn out to be a vice rather than a virtue. e.g.\ in case models with closed timelike curves turn out to be physically significant.\footnote{As is argued in \cite{earsmewut} and in \citet{smewut11}, this is a possibility that should not be ignored at the present stage of inquiry, given that it is difficult to know antecedently which parts of a new theory reveal important new physics.} That causal set theory cannot lend itself to spacetimes with closed timelike curves is not, however, a foregone conclusion: it might be that causal sets, although free of causal loops at the fundamental level, nevertheless can combine in ways such that at higher levels, causal loops emerge. If this turns out to be the case, however, then the emergent structure must necessarily violate the strictures of causal set theory and thus cannot be a model of it. I will return to this possibility in \S\ref{sec:emergent} below.

\subsection{Loop quantum gravity}\label{ssec:lqg}

Just like causal set theory, loop quantum gravity also starts from GR in its attempt to articulate a quantum theory of gravity. Instead of attempting this via the formulation of a classical discrete structure, it applies a canonical quantization procedure to GR. A canonical quantization of a classical theory attempts to preserve the core structure of the classical theory and convert it, in the most faithful way possible, into a quantum theory. This core structure consists in the canonical variables and their algebraic structure expressed by their Poisson bracket. The classical variables, such as position and momentum, are turned into quantum operators on a Hilbert space and the Poisson bracket becomes the commutation relation between the basic canonical operators. 

Any canonical approach to quantum gravity assumes that spacetime $\langle M, g_{ab}\rangle$ is globally hyperbolic and thus of topology $\Sigma\times\mathbb{R}$, where $\Sigma$ is again a three-dimensional spacelike submanifold of $M$.\footnote{An accessible textbook for both approaches to canonical quantum gravity described in this section is \citet{gampul}.} In this case (of global hyperbolicity), there exists a timelike vector field $t^a$ everywhere on $M$. This vector is tangent to a family of curves which can be parametrized by a `time' parameter $t$. The resulting three-dimensional surfaces $\Sigma_t$ of constant $t$ are totally ordered time slices in those spacetimes. 

An important technical choice for any canonical approach to quantum gravity is to select a pair of canonical variables as coordinates in the classical phase space of GR. In the traditional canonical approach, the four-dimensional metric of spacetime is rewritten as a function of the spatial three-metric defined on the $\Sigma_t$ and of the `lapse' $N$ and the `shift vector' $N^a$ resulting from the decomposition of $t^a = N n^a + N^a$, where $n^a$ is a vector field normal to the $\Sigma_t$'s. The pair of canonical variables in this approach is then given by the spatial three-metric as the `configuration' variable and what is essentially the extrinsic curvature as its conjugate momentum variable. 

Capturing the content of the globally hyperbolic sector of GR using this choice of canonical variables leads to a representational surplus resulting from expressing the physical content of the theory using more variables than are needed to capture the true degrees of freedom. As a consequence, `constraint equations' arise. The symmetric three-metric encodes six configuration degrees of freedom. The four constraint equations then leave us with two degrees of freedom for each point in space, as expected from GR. Solving the constraint equations thus gives us the true physical state space. Although its canonical variables permit a natural geometric interpretation, this choice is marred with insurmountable technical difficulties: the constraint equations are non-polynomial and no technique is known for solving them. This problem has essentially halted progress along these lines.

An alternative choice of basic variables promised to revive the canonical quantization programme and led to the approach known as `loop quantum gravity'. In loop quantum gravity, one proceeds not by using `metrical' variables to capture the geometry of spacetime, but instead variables based on the `connection'. Rewriting the metric in terms of `triads', the connection enters their covariant derivative, yielding an expression of the geometry of spacetime equivalent to that based on metrical variables. Loop quantum gravity then selects the so-called `Ashtekar variables', i.e., the (densitized) triads as momentum variables and the connection as canonically conjugate configuration variables. Re-expressing the Einstein-Hilbert action in terms of the components of the connection and the triads, it turns out that there are three sets of constraint equations which must be satisfied in order for the rewritten theory to be equivalent to (the globally hyperbolic sector of) GR. Among these, the Hamiltonian constraint is of particular interest and will be discussed in a moment. Moving to the quantum theory by means of canonical quantization, it seems natural to consider the `connection representation' of the wave function, i.e., expressing the wave function of the system as a function of the connection variable, similarly to Maxwell and Yang-Mills theories. However, technical difficulties suggest replacing the connection representation with the `loop representation', in which the wave function is given as a functional of `holonomies' around closed loops.\footnote{See \citet[Ch.\ 8]{gampul}.} 

Working in this loop representation renders two of the three families of constraint equations solvable. With just one constraint remaining to be solved, we arrive at what is known as the `kinematical Hilbert space'. The so-called `spin network states' can be constructed from the loop states and constitute an orthonormal basis of this Hilbert space \citep[\S6.3]{rov08lrr}. A spin network can be represented by a `coloured' graph such that both its nodes and the links between them carry spin representations. The spin network states can naturally be interpreted as forming a kind of discrete space where the nodes of the network represent the `atoms' of this granular space, and the links the surfaces where adjacent atoms `touch' \citep[\S1.2.2]{rov04}. On this interpretation, physical space is, fundamentally, a quantum superposition of spin networks.\footnote{For a further discussion concerning the physical interpretation of these spin networks, see \citet[\S2.1]{wut17}.} Although the quantum measurement problem prohibits a straightforward interpretation of this structure as chunky space, the geometric properties of the spin networks are at least suggestive of this natural interpretation. 

Time, on the other hand, seems to have disappeared entirely in canonical quantum gravity. The remaining constraint equation to be solved turns out to demand that the Hamiltonian operator sends the physical states to zero. Unlike in quantum mechanics, where the Schr\"odinger equation mandates how the Hamiltonian governs the dynamical evolution of the system, the Hamiltonian constraint equation here suggests that there is no change over time for genuinely physical states. In fact, there remains no quantity that could reasonably be interpreted as time in the Hamiltonian constraint equation.\footnote{\citet[\S2]{hugeal13} offers a more detailed explanation of the problem and brief survey of reactions to this `problem of time'.} Furthermore, this equation has so far resisted being solved, stalling the programme of loop quantum gravity. Without progress on this problem, however, we seem to have no prayer of even articulating what time travel could mean in this theory.

There are two workarounds. First, some physicists have symmetry-reduced the physical system under study, restricting the classical theory to homogeneous and isotropic spaces before subjecting it to quantization. This `cosmological sector' is much simpler than the full theory such that the corresponding Hamiltonian constraint equation can be solved. Unfortunately, these systems are too simple to permit anything that could reasonably be interpreted as time travel.\footnote{Though they are philosophically rich in other ways \citep{hugwut18}.} 

The second workaround is more relevant for our present purposes. The idea here is to forego the canonical description of the dynamical evolution in favour of a covariant formulation of the evolution. Hence, instead of the Hamiltonian operator, we express the dynamics of the theory in terms of transition amplitudes between `initial' and `final' kinematical states. These transition amplitudes are computed as weighted sums over `histories', i.e., ways in which the theory says the `final' state could have been obtained from the `initial' state. The details of how this is accomplished are irrelevant for our purposes (and are given in \citealt{rovvid}). What matters is that on a natural, but arguably overly simplistic, interpretation, both the `initial' and the `final' states deserve to be unquoted and correspond to quantum states of spatial hypersurfaces---indeed of global time slices of spacetime. Thus, we seem to be faced with a temporally innocuous structure in which no meaningful sense of time travel is permitted.

This interpretation is supported by the fact that any canonical quantization scheme of GR starts out by restricting itself to globally hyperbolic spacetimes. The canonical quantization recipe simply requires the classical spacetime structure of the physical system to be quantized to be globally hyperbolic, and thus causally well behaved. Just as for causal set theory, loop quantum gravity really only considers the globally hyperbolic sector of GR. One would therefore naturally assume that the theory also prohibits an analogue of closed timelike curves at the fundamental level, as did causal set theory. This conclusion would be premature, though. First, even though macroscopically the ordering of the initial and final states at earlier and later global times precludes closed curves in time, it could be that there exist tiny loops like this at the microscopic level. Second, just like causal sets, the spin networks of loop quantum gravity may combine such that causal loops emerge at a higher level even though there are none at the fundamental level. This second option will be discussed in \S\ref{sec:emergent} below, so let me finish with a brief word on the first possibility. 

Given that the problem of time has so far resisted resolution in the canonical approach to solving the Hamiltonian constraint of the full theory, the possibility of microscopic causality violations remains undecidable on this approach. On the covariant alternative, the possibility can be ruled out: the transition amplitudes are constructed from oriented `simplices' which are constructed from considering, among other things, the action of the Hamiltonian on the nodes of the spin network \citep[\S\S4.4, 5.3, 7.3]{rovvid}. Thus, the distinction between `timelike' and `spacelike' directions is maintained at the fundamental level. Given the construction rules of these simplices, microscopic causal loops are ruled out.

\subsection{String theory}\label{ssec:st}

As a third example of full quantum gravity, let us consider the fate of causal loops in string theory \citep{pol98,zwi04}. String theory is the dominant approach to quantum gravity. Unlike causal set theory and loop quantum gravity, it starts out from the standard model of particle physics and tries to extend the framework to incorporate gravity. As string theory is based on the paradigm of particle physics, it does not conceive of gravity as a feature of a dynamical spacetime, but instead as arising from an exchange of force particles, so-called `gravitons'. Furthermore, the point particles of earlier theories are replaced by 1-dimensional `strings' (or higher-dimensional `branes') in order to circumvent the problem of `non-renormalizability', which befell earlier attempts to incorporate gravity into the framework of particle physics \citep{wit96}. 

String theory exists at two levels. First, there is the perturbative level. At this level, string theorists have developed mathematical tools in order to define the string perturbative expansion over a given background spacetime. Second, the perturbative level is expected to be grounded in the more fundamental non-perturbative theory. This elusive `M-theory' does not yet exist. In fact, its existence is just inferred from the usual assumption that a perturbative expansion only ever gives an approximation to the true physical situation, which must be precisely captured by a more fundamental, and non-perturbative, theory. M-theory is thought to relate five different perturbative string theories by `dualities', i.e., symmetries equating strong coupling limits in one string theory to a weak coupling limit in another string theory. As M-theory does not yet exist, it is impossible to determine its verdict on time travel. 

However, supersymmetric gravity---widely considered a stepping stone towards full string theory---offers guidance into whether we should expect string theory to permit time travel. Although most of the results I am aware of have been obtained in five-dimensional supersymmetric gravity rather than in higher-dimensional theories, it turns out that solutions of five-dimensional supergravity can straightforwardly be extended to solutions of ten- or eleven-dimensional supergravity \citep[4590]{gaueal03}. Consequently, it appears as if string theory will likely admit time travel in case five-dimensional supersymmetric gravity does. And it turns out that five-dimensional supersymmetric gravity admits many solutions with closed timelike curves.

The systematic investigation of closed timelike curves in supersymmetric gravity starts two decades ago in \citealt{gibher99}. Since then, at least three important classes of solutions in five-dimensional supersymmetric gravity which contain closed timelike curves have been identified. First, there are supersymmetric solutions of flat space with a periodically identified time coordinate, resulting in a construction analogous to a rolled-up Minkowski spacetime of topology $S\times \mathbb{R}^3$ in GR. These solutions are topologically not simply-connected. In this case, passing to a covering spacetime avoids the closed timelike curves. In general, however, the supersymmetric solutions with closed timelike curves have a simply-connected topology and so cannot be avoided. 

The second class of supersymmetric solutions with closed timelike curves consists in an analogue of G\"odel spacetime \citep{gaueal03}. Just as G\"odel spacetime, these solutions model a topologically trivial, rotating, and homogeneous (and so not asymptotically flat) universe containing close timelike curves. Whether these G\"odel-type solutions really permit time travel has been contested: holography may effectively act to protect the chronology of G\"odel-type solutions in that closed timelike curves are either hidden behind a `holographic screen' and thus made inaccessible for timelike observers, or else broken up into pieces such that no closed timelike curves remain intact \citep{boyeal03}. 

The third class are the so-called `BMPV black hole' solutions, named after the initials of \citet{breeal97}. BMPV black holes are charged, rotating black holes in simply-connected, asymptotically flat spacetime. Thus, they are the supersymmetric counterparts of the general-relativistic Kerr-Newman black holes. Just as Kerr-Newman spacetimes can be maximally analytically extended to encompass a region inside the event horizon of the black hole to contain closed timelike curves \citep{wut99}, so can BMPV black holes, as has been shown by \citet{gibher99}. More precisely, Gibbons and Herdeiro show this to be the case for \textit{extremal} black holes, i.e., black holes whose angular momentum equals their mass (in natural units). It is unclear whether their result generalizes to include physically more realistic cases. Although they firmly establish their result only for a rather finely tuned combination of black hole parameters, \citet{gibher99} show that the presence of closed timelike curves for BMPV black holes is rather robust: this hyper-critical solution represents a simply connected, geodesically complete, asymptotically flat, non-singular, time-orientable, supersymmetric spacetime with finite mass, satisfying the dominant energy condition. Thus, `cosmic censorship', whatever its details, will struggle to eliminate this case.\footnote{See \citet[623]{smewut11} for more details.} 

None of these three classes of supersymmetric spacetimes with closed timelike curves conclusively establishes the possibility of time travel in supersymmetric gravity, let alone string theory. Having said that, however, it should be noted, with \citet{gaueal03}, that closed timelike curves appear generically in physically important classes of five-dimensional supersymmetric spacetimes. \citet{gaueal03} even complain how difficult it is to find five-dimensional solutions of supersymmetric gravity which do \textit{not} contain either closed timelike curves or singularities. 

Of course, this finding may be counted as a strike against supersymmetric gravity, rather than as a point in favour of time travel. Nevertheless, the emerging picture is one pointing toward the suggestion that closed timelike curves arise naturally in string theory, or at least in its vassal theories. Clearly, this suggestion remains preliminary in that it is wide open to what extent these results translate into a fundamental, non-perturbative version of string theory, and indeed whether string theory or any of the other approaches presented in this section are viable approaches to quantum gravity for that matter.

\section{Emergent time travel?}\label{sec:emergent}

In the last section, we have discussed the possibility that theories beyond GR directly issue a verdict on the permissibility of time travel. However, as stated in \S\ref{sec:intro}, we need to consider a second possibility, according to which an effective theory renders time travel physically possible, even though it is a valid approximation to a more fundamental theory, which in itself rules out time travel. This is the topic of this section. 

Of the four approaches discussed in \S\ref{sec:qg}, two seem to directly permit closed timelike curves and so time travel: while this was conjectured to be the case for string theory based on incomplete results from five-dimensional supersymmetric gravity, the prospects of some form of chronology protection obtaining are rather remote for semi-classical quantum gravity. Leaving aside the case of semi-classical quantum gravity, a note of caution concerning string theory. The results noted in the previous section pertain to the spacetime structure of `target space', which is the spacetime background for strings;\footnote{Strictly speaking, it is not even target space, or at least not the metric $g$ in it, is fundamental; rather, given a general metric in the action of a theory, one obtains a quantum theory of perturbations around a coherent state, which corresponds to the classical relativistic metric \citep[\S3]{hugvis15}.} it does not correspond to observed, `phenomenal' spacetime, which is an emergent phenomenon in string theory \citep{hug17}. If this is right, then regardless of whether the target spacetime contains closed timelike curves, what will be of interest is whether the emergent phenomenal spacetime will have a structure such as to permit time travel. As the emergence of spacetime, and particularly of its global properties, is at present only very partially understood in string theory,\footnote{See \citet{hugwut}.} further analysis of this will be left for another day. 

What is the situation in the two approaches which ruled out closed timelike curves (or their analogues) at the fundamental level, viz., causal set theory and loop quantum gravity? In a way, the situation for both causal set theory and loop quantum gravity is similar: fundamentally, they prohibit the equivalent of closed causal curves and so rule out time travel, as we have seen in the previous section. However, depending on what the relationship between the fundamental theory and emergent spacetime may be in each case, we may find that the emergent, macroscopic spacetime structure permits time travel. A consideration of the precise role and ambit of the theory for each case is necessary in order to appreciate this point.

One can distinguish between the astrophysical and the cosmological ambit of GR. On the one hand, GR furnishes a theory of gravity applicable to individual stars, or `small' isolated systems consisting of stars and smaller bodies such as our solar system. As such, it can describe the orbits of planets around their central star, the gravitational collapse of a star, a black hole, the merger of two black holes and the gravitational waves emitted on the occasion, and similar astrophysical phenomena involving gravity. 

On the other hand, since gravity is the dominant interaction at large distances, GR also delivers a cosmological theory, i.e., a theory describing the large-scale structure of the cosmos in its entirety and throughout most of its history. This should not be confused with a `theory of everything', which it clearly need not be despite the fact that it describes our world at the largest distances and over the longest durations. Qua cosmological theory, GR still supplies the backbone of the current cosmological standard model in the form of the Friedmann-Lema\^{\i}tre-Robertson-Walker spacetimes. 

These two applications of GR are---though connected---nevertheless distinct. Relativistic spacetimes describing phenomena of the astrophysical kind, just as the cosmological models, are often `global' or `large-scale' in that they encompass large (typically infinite) spatial distances and temporal durations as well. For instance, a Schwarzschild black hole is represented by a spacetime of infinite extent. However, such an astrophysical spacetime is not thought to correctly describe the large-scale structure of the cosmos at all: its description is accurate only near the astrophysical object it is thought to capture. The demand that such astrophysical spacetimes be asymptotically flat---roughly that the curvature vanishes away from the astrophysical object, i.e., in the `asymptotic' region---encodes the idea that the system at stake is, at least to a good approximation, isolated from the influence of other systems or indeed the rest of the cosmos.\footnote{To articulate precisely what asymptotic flatness amounts to, and, connectedly, what it is for a system to be isolated in a background-independent theory such as GR is far from trivial and requires some unpacking, as it is offered, e.g., in \citet[\S11]{wal84}.} In principle, spacetimes representing individual systems could then be stitched together in order to obtain a more complete description of the physics of ever larger and more encompassing parts of the cosmos. 

Closed timelike curves arise in both types of relativistic spacetimes. Astrophysical spacetimes such as Kerr-Newman spacetimes may contain causality-violating regions (in this case inside the maximal analytic extension of the interior of the black hole). For these spacetimes, closed timelike curves are typically confined to a region of spacetime. Thus, in general, there are events such that no closed timelike curves pass through them. Cosmological solutions such as G\"odel spacetime also accommodate time travel. In those cases, closed timelike curves are sometimes not confined to a region and thus in general every event lies on some closed timelike curves. in those cases, the opportunity to time travel is thus democratically awarded to all events. 

Returning to quantum gravity, if causal set theory and loop quantum gravity are considered cosmological theories, their laws reign supreme and one would not expect the possibility of time travel to arise. Indeed, since such cosmological models would have be consistent with the causality-enforcing features of these theories, the possibility of time travel would in this case be precluded universally. Let us consider this case for both theories separately.

\subsection{Causal set theory as a cosmological theory}
\label{ssec:causalcosmo}

Turning to causal set theory first, if the fundamental structure thus covers the cosmos and this structure is a causal set, then the condition of asymmetry entails that there cannot be a causal loop anywhere in the entire cosmos. Would it be possible, however, that even though causal loops are globally ruled out at the level of the fundamental structure the relativistic spacetime that emerges from this fundamental causal set contained closed timelike curves? 

In order to answer this question, we need to consider how relativistic spacetimes are thought to emerge from causal sets. While the emergence of spacetime in causal set theory has so far resisted resolution, the outlines are sufficiently clear for us to be in a position to settle the question.\footnote{For a much more detailed account of the emergence of spacetime in causal set theory, see \citet[Chs.\ 3, 4]{hugwut}.} As a necessary condition on the relationship between the underlying causal set and the emergent spacetime, there exists an embedding of the causal set into the spacetime. An embedding of a causal set into a spacetime is an injective map from the domain of the elements of the causal set into the manifold of the spacetime that preserves the causal structure in the sense that for any two elements $x$ and $y$ of the causal set, $x$ causally precedes $y$ if and only if the image of $x$ is contained in the causal past of the image of $y$. This condition and the asymmetry of the causal set together entail that any spacetime events in the image of the causal set cannot be part of a closed timelike curve.

Now it is consistent with the condition (and with the asymmetry of the causal set) that the emergent spacetime nevertheless contains closed timelike curves. If so, however, at most one of the events on the closed timelike curve could be in the image of the elements of the causal set. Thus, if there exist closed timelike curves in the emergent spacetime, there could be absolutely no trace of this fact in the fundamental structure. As a causal set is discrete and a relativistic spacetime a continuum structure, it will in general not be the case that the fundamental causal set contains all the `information' present in a relativistic spacetime. That the emergent spacetime not contain any relevant geometric features not already in some form present in the causal set motivates the additional demand that the embedding be \textit{faithful}, i.e., that the map distributes the images of the elements of the causal set approximately uniformly on the spacetime manifold, which is assumed to be approximately flat below the discreteness scale.\footnote{For a more precise formulation, see \citet[Ch.\ 4]{hugwut}.} The idea behind imposing faithfulness is precisely that the geometry of the emergent spacetime be `boring' below the scale captured (and \textit{capturable}) by the fundamental discrete structure. 

If the emergent spacetime contained---presumably at Planckian scales---very thin slices disconnecting from the bulk of the spacetime, looping back to reconnect to it at earlier times in a way such that it contained closed timelike curves running along these slices, then it may not violate faithfulness: the spacetime could be flat (locally Minkowskian) everywhere with just no image point on the thin slice looping back. Though thus consistent with the letter of faithfulness, such a situation would arguably violate its spirit: that the emergent spacetime not contain any relevant features not at least implicitly present in the fundamental structure. In sum, if causal set theory is regarded as a cosmological theory, there appears to be quite literally little space for an emergent spacetime to naturally accommodate closed timelike curves.

\subsection{Loop quantum gravity as a cosmological theory}
\label{ssec:loopcosmo}

Many of the conclusions arrived at in the case of causal set theory qua cosmological theory hold in the case of loop quantum gravity in this regime as well. In fact, there is explicit consideration of a sector of loop quantum gravity, known as `loop quantum cosmology', which studies symmetry-reduced models of loop quantum gravity. By imposing isotropy and homogeneity already at the classical level, the constraint equations simplify sufficiently to admit explicit solutions \citep{boj11}. In those models, a `cosmic' time totally orders all events and there is consequently no possibility for time travel. 

More generally, the Hamiltonian formalism presupposes the physical system at stake---be it a pendulum, a planet, or spacetime itself---is spatially extended and evolves over time, following the dynamics of Hamilton's equation. Classically, this assumes, as we saw above, that the spacetime has topology $\Sigma\times \mathbb{R}$ such that the spatial time slices $\Sigma$ are again totally ordered in `time' by the reals. Moving to the quantum theory, as (to repeat) the canonical programme has stalled, the question remains open what the states in the physical Hilbert space are, and so how they ought to be interpreted physically. 

Alternatively, covariant loop quantum gravity does not easily lend itself to a cosmological interpretation. The `initial' and `final time' slices are intended as such, and the spacetime region they enfold is finite and generally rather small. Independently of the size of the region enveloped by the time slices, a truly cosmological model cannot in general be expected to have a first or last `moment' in time. Although one could in principle identify the initial and final time slices and so create a model with the equivalent of closed timelike curves, such constructions would be an abuse of the theory clearly beyond its intended ambit. Thus, fundamentally, cosmological loop quantum gravity does not permit time travel.

Could there be time travel in emergent spacetime, perhaps by means of some more or less artificial construction? Unfortunately, this cannot be conclusively answered, since the emergence of spacetime from states in loop quantum gravity is yet to be fully understood.\footnote{See \citet{wut17} for a more detailed sketch of the current state of the art.} Although the possibility of finely carved emergent spacetimes with closed timelike curves cannot be excluded, it seems as if such spacetimes should not emerge from full-sized cosmological states in loop quantum gravity.

\subsection{Quantum gravity as `astrophysics'}
\label{ssec:qgastro}

There is an alternative to considering a quantum theory of gravity as offering a cosmological theory: it may be deemed, rather, as describing much more local phenomena, such as astrophysical black holes or the very early universe.\footnote{The latter is of course not really a `local' phenomenon as it concerns the early stages of the whole cosmos; however, since the description is really of a very small universe during the first few `Planck times', the description would be only of what is really a very small part of spacetime. This is indeed the remit of `quantum cosmology', which thus becomes an `astrophysical' theory under the present use of the term.} In fact, these are the phenomena where most physicists expect that only a quantum theory of gravity could deliver a satisfactory account, motivating quantum gravity in the first place. 

Although there are some efforts in this direction (such as the estimation of an entropy bound in \citealt{ridzoh06}), causal set theory lacks well-developed astrophysical applications. This is largely owed to the fact that it is to date a classical theory whose transformation to a quantum theory has been but roughly sketched. Apart from the cosmological applications mentioned in \S\ref{ssec:loopcosmo}, loop quantum gravity has also seen some research on black holes (such as the derivation of an expression for the black hole entropy similar to the usual Bekenstein-Hawking formula in \citealt{rov96} or studies of black hole singularity resolution e.g.\ in \citealt{gampul13}). As far as I can tell, there is no indication of the possibility of time travel in any of these applications. 

\begin{figure}
\centering
\epsfig{figure=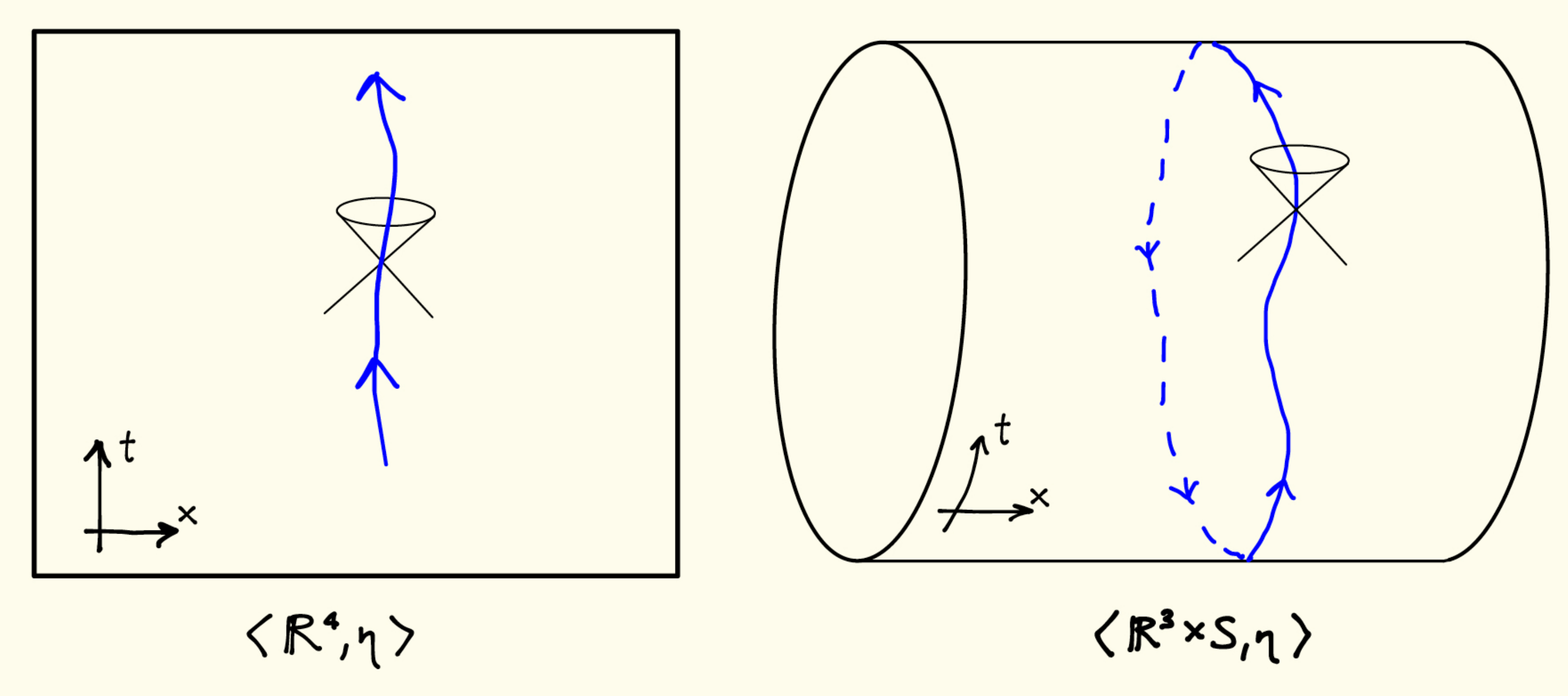,scale=0.46}
\caption{\label{fig:twospts} Two locally isometric spacetimes, only one of which contains closed timelike curves.}
\end{figure}

But there remains another option. Perhaps quantum gravity will ever only be concerned with local phenomena, offering a fundamental description of the finest threads of what is spacetime macroscopically, while never amounting to a theory of global structure. If so, a quantum theory of gravity should not be considered cosmological. Instead, the global structure would emerge from patching together smaller pieces of fundamental quantum gravitational structures to cosmological totalities following principles or laws distinct from those asserted in quantum gravity. In fact, in GR itself, we cannot infer from a locally causally well behaved spacetime that it contains no closed timelike curves and so is globally well behaved. There exist pairs of locally isometric spacetimes such that one of them contains closed timelike curves while the other does not. For instance, Minkowski spacetime $\langle \mathbb{R}^4, \eta\rangle$ and a slice of Minkowksi spacetime rolled up along the timelike direction are locally isometric and so physically indistinguishable, as is illustrated in Figure \ref{fig:twospts}.

Whether or not time travel remains possible in those constructs thus depends on the nature of these laws governing the global structure. If they are as permissive as those in GR (or indeed \textit{are} those of GR), then the resulting global structure will admit (whatever corresponds to) closed timelike curves and time travel in this sense is possible. Of course, these laws may also be more restrictive and preclude the possibility of time travel. For now, the question remains wide open.

\section{Conclusions}\label{sec:conc}

One may hope, with \citet*{andnemwut}, that a theory more fundamental than GR would deliver insight into the physical mechanism (such as rotation or `antirotation') behind `acausalities' arising in GR such as the presence of closed timelike curves in some relativistic spacetimes. This hope may be disappointed, even though a more fundamental theory may well admit structures amounting to closed timelike curves and thus permit time travel. As the deliberations in this article show, this clearly remains a live option at the present stage of knowledge. Unfortunately, it is also presently impossible to pronounce any even tentatively conclusive lessons concerning the possibility of time travel to be drawn from quantum gravity. Any more definite insight must await a fuller development of the field. 

In fact, the preliminary analysis above illustrates just how little we currently know regarding the relationship between these more fundamental theories of quantum gravity and GR. While a fuller analysis of the relationship between quantum gravity and GR is beyond the scope of the present article,\footnote{\citet{hugwut} consider the state of the art regarding the relationship between quantum theories of gravity and GR much more fully.} the issue of what can be said about the causal structure of spacetime as a `classical' limit of the underlying theories of quantum gravity in general, and about the emergence of closed timelike curves in particular, exemplifies that much work remains to be done in quantum gravity.\footnote{I thank the anonymous referee for pressing this conclusion. I agree that this is an important upshot of my discussion.}

Formulated more positively, although we have yet to learn whether time travel is possible or not, our study  blazes a trail forward: using the possibility of time travel and its attendant consideration of the causal structure of spacetime as our foil, the above analysis has led us into the heart of the nature of quantum gravity, its ambit, and---centrally---its relation to GR. For this reason alone, the question of time travel beyond GR is worth our while. Even as we await more determinate answers.

\bibliographystyle{plainnat}
\bibliography{/Users/christian/Professional/Bibliographies/timetravel}

\end{document}